\documentclass[twocolumn, showpacs,preprintnumbers,amsmath,amssymb]{revtex4}
\usepackage{graphicx}% Include figure files
\usepackage{dcolumn}% Align table columns on decimal point
\usepackage{bm}% bold math
\begin{document}

\preprint{APS/123-QED}

\title{Stopping power: Effect of the projectile deceleration}% Force line breaks with \\

\author{Roman Kompaneets}
\email{kompaneets@mpe.mpg.de}
\affiliation{Max-Planck-Institut f\"ur extraterrestrische Physik, Giessenbachstr. 1, 85748 Garching, Germany}

\author{Alexei V. Ivlev}
\affiliation{Max-Planck-Institut f\"ur extraterrestrische Physik, Giessenbachstr. 1, 85748 Garching, Germany}

\author{Gregor E. Morfill}
\affiliation{Max-Planck-Institut f\"ur extraterrestrische Physik, Giessenbachstr. 1, 85748 Garching, Germany}

\date{\today}% It is always \today, today,

\begin{abstract}
The stopping force is the force exerted on a charged projectile
by the excess charge of the wake generated by the projectile in the surrounding plasma.
Since the wake does not instantly adjust to the projectile velocity, 
the stopping force should be affected by the projectile deceleration
caused by the stopping force itself. We address this effect by 
deriving the corresponding correction to the stopping force in a cold plasma. We find that if the projectile 
is an ion, the correction is small when the stopping force is due to the plasma electrons,
but can be significant when the stopping force is due to the ions.
\end{abstract}

%\keywords{...}%Use showkeys class option if keyword
\maketitle

\section{Introduction}
\label{introduction}
The calculation of the stopping force, which is the force exerted by a moving charged projectile
on itself through the perturbation of the surrounding plasma, is a classic problem,
with applications spanning from fusion~\cite{peter-mey-phys.rev.a-1991, hoffmann, kawata-2007}
to dusty plasmas~\cite{lampe-joy-gan-phys.plasmas-2000, khrapak-phys.rev.e-2002, yaroshenko-2005, ishihara-j.phys.d.appl.phys-2007, miloch}. 
In the linear perturbation 
approximation, 
the calculation of the stopping force is straightforward
once the dielectric function is specified and the projectile velocity is assumed to be 
constant~\cite{peter-mey-phys.rev.a-1991, ivlev-zhd-khr-phys.rev.e-2005};
for a cold plasma, it results in the Bohr 
stopping force~\cite{bohr, peter-mey-phys.rev.a-1991}.

While the standard calculation assumes a constant projectile velocity,
in reality projectiles experience deceleration for the stopping force itself.
Since
the plasma perturbation, or wake, does not instantly adjust to the projectile velocity,
the momentary stopping force should be different from the one predicted by the standard calculation using the momentary value of the velocity.

This difference, that is, the effect of the deceleration on the stopping force, is not obvious to estimate. 
For a cold plasma and uniform motion, as well known, one must include a
damping rate to avoid a singularity of the integral for the stopping force~\cite{peter-j.plasma.phys-1990, kompaneets-vla-ivl-new.j.phys-2008}; 
the inclusion of a deceleration, as we will show, results in a competition
between the deceleration
and the damping [Eq.~(\ref{line-contribution})].
When the deceleration dominates, which is a quite common regime [as shown in Sec.~\ref{discussion}], 
it is difficult to make simple estimates on the effect of the deceleration on the stopping force
without making accurate calculations.

Given the universality of the effect and that its magnitude is not obvious, it needs to be rigorously addressed,
which is the object of this paper.

\section{Theory}

Let us consider a point charge $q$ moving through a plasma
according to
\begin{equation}
x(t)=y(t)=0, \quad z(t)=\frac{at^2}{2}
\end{equation}
with a constant $a>0$ starting from $t=-\infty$
and derive the stopping force on that charge at any $t<0$,
the decelerating phase of the motion,
first assuming 
an arbitrary dielectric function
$D(\omega, {\bf k})$ and then focusing on the case of 
a cold plasma.
Note that by making the assumption of a constant deceleration, we probe into the principal 
effect of non-uniform motion and neglect the higher-order corrections.

For arbitrary $D(\omega, {\bf k})$, the steps to derive 
an integral expression for the stopping force
are rather straightforward.
The extraneous charge density
in our case is
\begin{equation}
\rho({\bf r},t)= q \delta(x) \delta (y) \delta \left( z-\frac{at^2}{2}\right),
\end{equation}
so its Fourier transform in space and time is
\begin{eqnarray}
\hat{\rho}(\omega, {\bf k})=\int_{-\infty}^{\infty} dt \, \int d{\bf r} \, \rho({\bf r},t) \exp(i\omega t - i{\bf k} \cdot {\bf r})
\nonumber \\
=q\sqrt{\frac{\pi}{|k_z| a}} \exp\left(  \frac{i \omega^2}{2k_z a} \right) \left(1-\frac{ik_z}{|k_z|}\right).
\end{eqnarray}
We use the linear perturbation approximation,
\begin{equation}
\hat \varphi_{\rm diff}(\omega, {\bf k})=\frac{\hat \rho(\omega,{\bf k})}{\varepsilon_0 k^2} \left( \frac{1}{D(\omega, {\bf k})}-1 \right),
\label{linear-perturbation}
\end{equation}
where $\hat \varphi_{\rm diff}(\omega, {\bf k})$ is the Fourier transform
in space and time of the difference $\varphi_{\rm diff}({\bf r}, t)$ between the potential induced by the charge $q$
and its Coulomb potential, 
and $\varepsilon_0$ is the permittivity of free space.
The stopping force
\begin{equation}
F_{\rm  }(t)=-q \frac{\partial \varphi_{\rm diff}({\bf r}, t)}{\partial z},
\end{equation}
where the derivative is taken at the charge location,
is found by performing the inverse Fourier transform of Eq.~(\ref{linear-perturbation}).
We simplify the resulting expression by (i) confining the integration to positive $k_z$ --- for this we use the general symmetry properties
${\rm Re} D(\omega, {\bf k})={\rm Re} D(-\omega, -{\bf k})$ and 
${\rm Im} D(\omega, {\bf k})=-{\rm Im} D(-\omega, -{\bf k})$ ---
and (ii) making the substitution
\begin{equation}
\omega=\eta \sqrt{k_z a} + k_z a t,
\end{equation}
where $\eta$ is a new integration variable.
The result is
\begin{eqnarray}
F_{}=
-\frac{q^2\sqrt{\pi}}{8\pi^4 \varepsilon_0} {\rm Re}
\int_{k_z>0} \frac{k_z \, d{\bf k}}{k^2}
\int_{-\infty}^{\infty} d\eta \, 
(1+i) \exp \left( \frac{i\eta^2}{2} \right) 
\nonumber \\
\times \left[
\frac{1}{D(k_z v_z + \eta\sqrt{k_z a}, {\bf k})}-1
\right],
\label{general-expression}
\end{eqnarray}
where $v_z=at$ is the momentary $z$-velocity of the charge.
Here and in the following, the integration over ${\bf k}$ is restricted
to $|k_z|<k_{\rm max}$ and $k_\perp(=\sqrt{k_x^2+k_y^2})< k_{\rm max}$,
where $k_{\rm max}$ is the inverse length at which the nonlinear effects
become significant, in order to avoid the well-known logarithmic divergence
at large wave numbers~\cite{peter-mey-phys.rev.a-1991, ivlev-zhd-khr-phys.rev.e-2005}.
(Note that we could alternatively restrict the integration by 
$k < k_{\rm max}$ as in Ref.~\cite{peter-mey-phys.rev.a-1991}; both approaches are valid when $k_{\rm max}$ is sufficiently large for the
linear perturbation approximation to be applicable.)

It is easy to see that for $a = 0$ Eq.~(\ref{general-expression}) reduces 
to the well-known expression for the stopping force on a uniformly moving charge for an arbitrary 
dielectric function ~\cite{ivlev-zhd-khr-phys.rev.e-2005}.

For a nonzero $a$, even if $a$ is assumed to be small, it seems impossible to substantially 
simplify Eq.~(\ref{general-expression}) unless $D(\omega, {\bf k})$ is specified. 
For instance, the series expansion of the integrand in Eq.~(\ref{general-expression}) in powers of $\sqrt{a}$ (with $v_z$ being an
independent parameter)
produces terms $\propto \exp(i\eta^2/2) \eta^{n} a^{n/2}$, integrals of which over $\eta$
diverge for non-zero $n$, while methods of complex variable integration, employed below, can be used only
when the analytical properties of $D(\omega, {\bf k})$ are known.

So let us focus on the case of a cold plasma, assuming~\cite{lifshitz-pitaevskii-1981, kompaneets-vla-ivl-new.j.phys-2008}
\begin{equation}
D(\omega, {\bf k})=1-\frac{\omega_{\rm pe}^2}{(\omega + i0^{+})^2},
\label{cold-plasma}
\end{equation}
where $\omega_{\rm pe}$ is the plasma frequency.
The term $i0^{+}$ (where $0^{+}$ is an infinitesimal positive number)
represents the collisional and Landau damping and is important as it removes
the singularity of the integral in Eq.~(\ref{general-expression})~\cite{peter-j.plasma.phys-1990, kompaneets-vla-ivl-new.j.phys-2008}.
We replace the term $i0^{+}$ by a finite imaginary number $i\nu$ 
and keep considering it finite for a while (before taking the limit $\nu \to 0$), which will allows us to 
make estimates on the role of finite damping. Note that for a cold plasma, the upper limit $k_{\rm max}$ is~\cite{peter-mey-phys.rev.a-1991}
\begin{equation}
k_{\rm max}=\frac{4 \pi \varepsilon_0 m_{\rm e} v_z^2}{|q|e},
\end{equation}
where $m_{\rm e}$ is the electron mass and $e$ is the elementary charge.

We substitute the dielectric function~(\ref{cold-plasma}) (with $i0^{+}$ replaced by $i\nu$) to 
Eq.~(\ref{general-expression})
and normalize the quantities as follows:
\begin{eqnarray}
F_{}   \left( \frac{q^2 \omega_{\rm pe}^2}{4 \pi \varepsilon_0 v_z^2}\right)^{-1} 
\to F_{}, \quad \frac{{\bf k} |v_z|}{\omega_{\rm pe}} \to {\bf k}, \nonumber \\
\frac{a}{|v_z| \omega_{\rm pe}} \to a, \quad \frac{\nu}{\omega_{\rm pe}} \to \nu, \quad \frac{k_{\rm max} |v_z|}{\omega_{\rm pe}} \to k_{\rm max},
\end{eqnarray}
which yields (for $v_z<0$)
\begin{eqnarray}
F_{}=-\frac{1}{2\pi^{5/2}} {\rm Re} \int_{k_z>0} \frac{k_z \, d{\bf k}}{k^2}
\nonumber \\
\times
\int_{-\infty}^{\infty} 
\frac{(1+i) \exp(i\eta^2/2) \, d\eta}{(k_z-\eta\sqrt{k_z a}-i\nu)^2-1}.
\label{before-contour}
\end{eqnarray}

\begin{figure}
\includegraphics[width=5 cm]{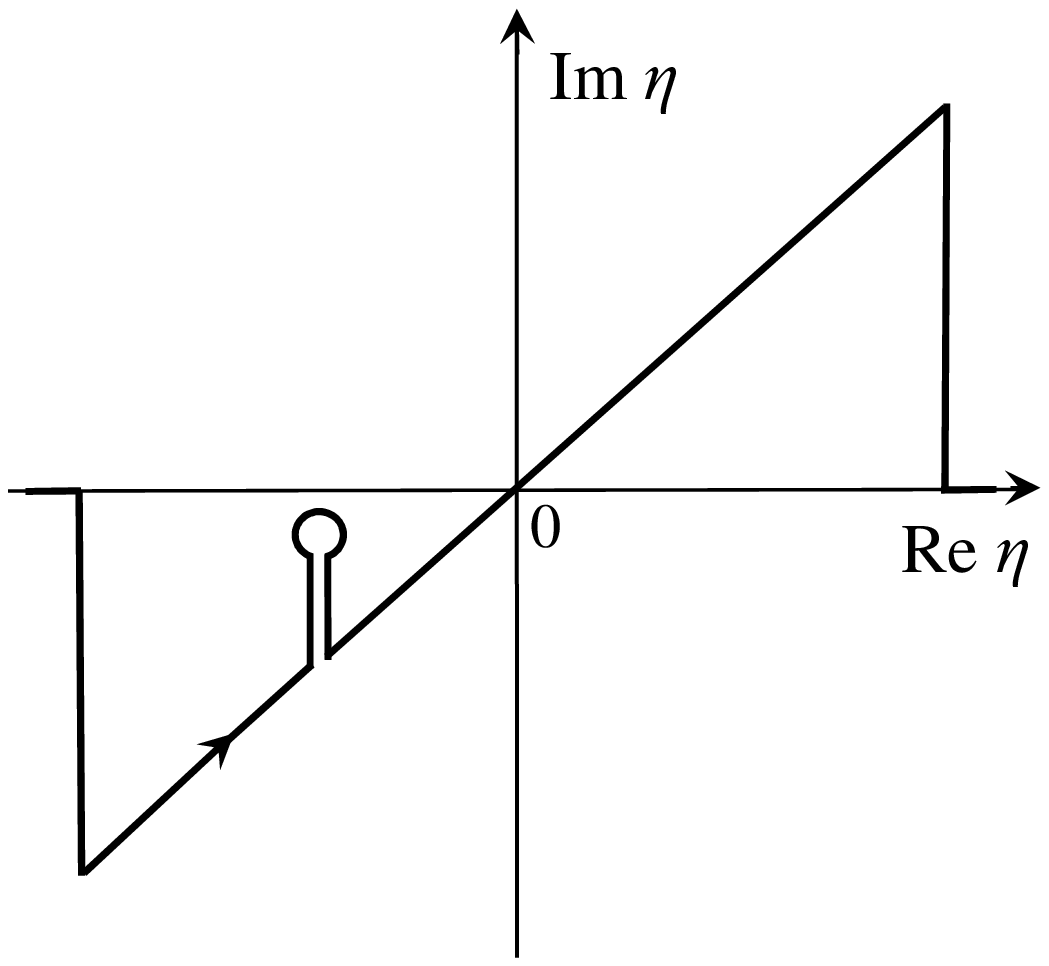}
\caption{The modified contour to calculate the integral over $\eta$ in Eq.~(\ref{before-contour}).
The long vertical lines at the left and right are located at infinitely large ${\rm Re}\, \eta$. The oblique line
is at $45$ degrees to the horizontal axis. The circle has an infinitely small radius
and is centered at the singularity $\eta=(-1+k_z-i\nu)/\sqrt{k_z a}$ if this point lies above
the oblique line (otherwise, the circle is not included).}
\label{sketch}
\end{figure}

As noted above, the formal expansion of the integrand in powers of $\sqrt{a}$ 
produces divergent integrals over $\eta$, which is because of a slow convergence of the original integral over $\eta$, 
so we need to transform it to improve
the convergence.
For this purpose, we 
modify the contour of integration as shown in Fig.~\ref{sketch} ---
according to Cauchy's theorem, this does not change the integration result.
It is easy to show that since we choose the distance from the vertical lines of the contour to the imaginary axis to be infinitely large,
the contributions from them are zero.
We get
\begin{equation}
F_{\rm  }=F_{\rm  }^{\rm pole}+ F_{\rm  }^{\rm line},
\label{sum-contr}
\end{equation}
where the contribution from the pole and the line are, respectively,
\begin{eqnarray}
F_{\rm  }^{\rm pole}=
\frac{1}{2\pi^{3/2}} {\rm Re}  \int_{0<k_z<1-\nu} \frac{d{\bf k}}{k^2} \,
(1-i)\sqrt{\frac{k_z}{a}} \nonumber \\
\times \exp \left[ \frac{i(k_z-1-i\nu)^2}{2k_z a} \right],
\label{pole-contribution}
\end{eqnarray}
and
\begin{eqnarray}
F_{}^{\rm line}=-\frac{1}{\pi^{5/2}} {\rm Re} \int_{k_z>0} \frac{ik_z \, d{\bf k}}{k^2} \, \int_{-\infty}^{\infty} d\xi \,
\nonumber \\
\times \frac{\exp \left( -\xi^2 \right)}{[k_z-\xi (1+i) \sqrt{k_z a}-i\nu]^2-1}.
\label{line-contribution}
\end{eqnarray}
[Here, we have made the substitution $\eta=(1+i) \xi$.]
Now the convergence is considerably improved as the integral over $\xi$
converges exponentially fast.

We note that when both $a$ and $\nu$ are infinitesimal, 
whether the singularity $k_z=1$ in Eq.~(\ref{line-contribution}) is passed above or below for a given $\xi$
depends on the sign of $\xi \sqrt{a}+\nu$. Since the main contribution
to the integral comes from $|\xi|\sim 1$, it is clear that
for $\sqrt{a} \ll \nu$
we essentially repeat
the standard derivation of the stopping force for a uniform motion
(in which the singularity is passed below). Mathematically, this corresponds 
to taking the limit $a \to 0$ first, then $\nu \to 0$.
In this case, we get $F_{\rm  }^{\rm pole}=0$ and $F_{{\rm }}^{\rm line}=(1/2)\ln(k_{\rm max}^2+1)\simeq \ln k_{\rm max}$ 
(for $k_{\rm max} \gg 1$,
which is the condition to employ the linear perturbation approximation),
which is the Bohr stopping force.
Furthermore, by using Eqs.~(\ref{pole-contribution}) and (\ref{line-contribution}), it is easy to show that
for $\sqrt{a} \ll \nu$ the correction due to a finite $a$ is small as compared to
the correction due to a finite $\nu$.

So let us focus on the limit $\nu \to 0$ for a finite $a$, which implies the opposite regime, $\sqrt{a} \gg \nu$.
To do so, it is sufficient to simply set $\nu=0$
in Eqs.~(\ref{pole-contribution}) and (\ref{line-contribution}) because a finite $a$ already removes the singularity in Eq.~(\ref{line-contribution});
note that we have already used that $\nu >0$
by including the contribution from the pole to Eq.~(\ref{sum-contr}).

To calculate $F_{\rm  }^{\rm pole}$, we first make the substitution $\beta = (1-k_z)^2/(a k_z)$, where $\beta$ is a new integration variable,
so the exponent
in Eq.~(\ref{pole-contribution})
becomes $\exp(i\beta/2)$. To improve the convergence, we change
the integration over $\beta$ from the real to imaginary axis, with the new integration limits being $0$ and $+i\infty$.
(Again, the result remains the same according to Cauchy's theorem.) Then we expand the integrand in powers
of $\sqrt{a}$ and integrate the resulting terms over $\beta$,
which yields
\begin{eqnarray}
F_{\rm  }^{\rm pole}= \frac{1}{2} \ln(k_{\rm max}^2+1) \nonumber \\
+ \frac{1}{2}\sqrt{ \frac{a}{\pi}} \int_0^{k_{\rm max}} \frac{(1-3k_\perp^2) k_\perp}
{(k_\perp^2+1)^2}\, d k_\perp + o(a)
\label{pole1}
\end{eqnarray}
for $a \to 0$.
Note that the term $\propto a$ turns out to be zero.
Also note that in the zeroth order, $F_{\rm  }^{\rm pole}$
is the Bohr stopping force (again, for $k_{\rm max} \gg 1$),
which is in contrast to the case where the limit $a \to 0$ is taken before the limit $\nu \to 0$.

The calculation of $F_{\rm  }^{\rm line}$
in the zeroth order is rather straightforward and yields zero, while
in order 
to calculate the $\sqrt{a}$-term for $F_{\rm  }^{\rm line}$,
we differentiate the integrand in Eq.~(\ref{line-contribution}) over $\sqrt{a}$
and integrate it over $\xi$ and ${\bf k}$ in the limit $a \to 0$. We consider 
the integral over $k_z$ first and
divide it into two parts,
one from the vicinity of $k_z=1$ and the other one from the rest of the integration interval over $k_z$.
The latter part obviously cancels out when subsequently integrated over $\xi$, while the former one
is easily found by evaluating the corresponding residue.
The subsequent integration over $\xi$ is straightforward.
The resulting $\sqrt{a}$-term turns out to be minus
the $\sqrt{a}$-term in Eq.~(\ref{pole1}), so they cancel out.

To calculate the next expansion term for $F_{\rm  }^{\rm line}$, we 
differentiate the integrand in Eq.~(\ref{line-contribution}) over $\sqrt{a}$ twice
and integrate it over $\xi$ and ${\bf k}$ in the limit $a \to 0$.
Since the limit $a \to 0$ is considered, after the differentiation we can replace
$\xi (1+i) \sqrt{k_z a}$ in the denominator by $i\varepsilon \xi/|\xi|$,
where $\varepsilon$ is an infinitesimal positive number (independent of $k_z$).
We consider the integration over $k_z$ first and expand the integration interval
from $0<k_z<k_{\rm max}$ to $-\infty<k_z<\infty$ as the integral converges well at large $k_z$
and the integrand is an even function of $k_z$. This modification allows us to integrate over $k_z$
by
using the residue theorem
and Jordan's lemma. The subsequent integration over $\xi$ and $k_\perp$ is straightforward --- again, since
the integral over $k_\perp$ converges well at large $k_\perp$,
we replace the upper 
limit $k_{\rm max}$ by $\infty$. The result for the expansion term turns out to be $-\pi a/2$.

By summing $F_{\rm  }^{\rm pole}$ and $F_{\rm  }^{\rm line}$ and replacing $(1/2)\ln (k_{\rm max}^2+1)$ by $\ln k_{\rm max}$, 
we can write
\begin{equation}
F=F_0+\delta F,
\label{principal1}
\end{equation}
where, in the dimensional units,
\begin{equation}
F_0 = \frac{q^2 \omega_{\rm pe}^2}{4 \pi \varepsilon_0 v_z^2} \ln \left( \frac{ k_{\rm max} |v_z|}{\omega_{\rm pe}} \right)
\label{bohr}
\end{equation}
is the Bohr stopping force and
\begin{equation}
\delta F_{\rm  }= -\frac{q^2 \omega_{\rm pe} a}{8 \varepsilon_0 |v_z|^3}
\label{principal2}
\end{equation}
is the deceleration correction, where we have omitted the higher-order terms with respect to $a$.

Equation~(\ref{principal2}) is the principal result of this paper.
The derived correction is negative, which means that the stopping force
is reduced as a result of the deceleration. 
We also note that the correction
does not contain the Coulomb logarithm.

\section{Discussion and conclusions}
\label{discussion}
To estimate whether and when the deceleration correction can be significant,
let us use that the deceleration is caused by
the stopping force itself. We assume that the projectile
is an ion of charge $Ze$ and mass $Am_{\rm p}$
and moves with a velocity $v \gtrsim 2 v_{\rm e}$,
the range where the Bohr expression for the stopping force is applicable~\cite{peter-mey-phys.rev.a-1991}.
Here, $m_{\rm p}$ is the proton mass,
$v_{\rm e}=\sqrt{k_{\rm B}T_{\rm e}/m_{\rm e}}$ is the electron thermal velocity,
$T_{\rm e}$ is the electron temperature, and $k_{\rm B}$ is the Boltzmann constant.
The deceleration is then taken to be $a=F_0/(Am_{\rm p})$.

First of all, we analyze when the condition $\sqrt{a} \gg \nu$ (in the dimensionless units) holds, i.e., when the deceleration
dominates over the damping, as assumed in the derivation of Eqs.~(\ref{principal1})-(\ref{principal2}). In the dimensional units, this condition
reads
\begin{equation}
\frac{a \omega_{\rm pe}}{v} \gg \nu^2.
\label{role-of-nu}
\end{equation}
We substitute one half of the effective electron-ion Coulomb collision frequency for $\nu$ (see Ref.~\cite{aleksandrov-book}),
\begin{equation}
\nu=\frac{1}{6 (2 \pi)^{3/2}} \frac{v_{\rm e} n e^4}{(\varepsilon_0 k_{\rm B} T_{\rm e})^2} \ln \Lambda_{\rm e},
\end{equation}
where the ions are assumed to be singly ionized, $n$ is the plasma number density,
\begin{equation}
\Lambda_{\rm e}=\frac{4\pi \varepsilon_0 k_{\rm B} T_{\rm e} v_{\rm e}}{\omega_{\rm pe}e^2},
\end{equation}
and $\omega_{\rm pe}=[n e^2 / (\varepsilon_0 m_{\rm e})]^{1/2}$. This allows us to rewrite Eq.~(\ref{role-of-nu}) as
\begin{equation}
\frac{1}{\Gamma_{\rm e}^{3/2}} \frac{Z^2}{A} \left( \frac{v_{\rm e}}{v} \right)^3 \frac{\ln \Lambda_q}{\ln^2 \Lambda_{\rm e}} \gg 10^{2},
\label{small-nu-estimate}
\end{equation}
where the factor $10^{2}$ includes all numerical coefficients as well as the proton-electron mass ratio,
\begin{equation}
\Gamma_{\rm e}=\frac{e^2 n^{1/3}}{4\pi \varepsilon_0 k_{\rm B} T_{\rm e}}
\end{equation}
is the electron coupling parameter, and $\Lambda_q$ is the argument of the logarithm in Eq.~(\ref{bohr}).
Clearly, the condition~(\ref{small-nu-estimate}) is often satisfied --- the electrons only need to be sufficiently weakly coupled for that.

The ratio of the deceleration correction~(\ref{principal2}) to the Bohr stopping force~(\ref{bohr}) can be written as
\begin{equation}
\frac{|\delta F|}{F_0} \simeq 3\times 10^{-3} \Gamma_{\rm e}^{3/2} \frac{Z^2}{A} \left( \frac{v_{\rm e}}{v} \right)^{3},
\label{ratio-electrons}
\end{equation}
where, again, the factor $3\times 10^{-3}$ includes all numerical coefficients and the electron-proton mass ratio.
We see that there is no way
of making this correction comparable to the Bohr stopping force as long as the electrons are weakly coupled
[as required by Eq.~(\ref{small-nu-estimate})] and $v \gtrsim 2 v_{\rm e}$ (as assumed above).

Let us now consider the regime
$2v_{\rm i} \lesssim v \ll v_{\rm e}$, where $v_{\rm i}=\sqrt{k_{\rm B}T_{\rm i}/m_{\rm i}}$ is the ion thermal velocity,
$T_{\rm i}$ is the ion temperature, and $m_{\rm i}$ is the ion mass.
In this case, we can still use 
Eqs.~(\ref{principal1})-(\ref{principal2}), but we must replace the quantities characterizing the electron component by 
those for the ion component. Then the ratio of the deceleration correction to the Bohr stopping force, which is now due to the ions,
is
\begin{equation}
\frac{|\delta F|}{F_0} =(\pi \Gamma_{\rm i})^{3/2} \frac{Z^2}{A} \left( \frac{v_{\rm i}}{v} \right)^{3},
\label{ratio-ions}
\end{equation}
where 
\begin{equation}
\Gamma_{\rm i}=\frac{e^2 n^{1/3}}{4\pi \varepsilon_0 k_{\rm B} T_{\rm i}}
\end{equation}
is the ion coupling parameter. [Note that the restriction~(\ref{small-nu-estimate}) is inapplicable here as it was derived for electrons;
for ions there is no collisional background unless the plasma is partially ionized.]
Equation~(\ref{ratio-ions}) shows that in the regime considered, 
the deceleration effect can be highly significant.
For instance, for
$v=2v_{\rm i}$ (which is approximately the location of the peak of the stopping force~\cite{peter-mey-phys.rev.a-1991}),
an S$^{16+}$-projectile (as, e.g., in the experiments of Ref.~\cite{betz-hop-sch-1988}), a hydrogen plasma,
and $\Gamma_{\rm i}=0.5$,
Eq.~(\ref{ratio-ions}) yields $\delta F/F_0 \simeq 2$, meaning that the peak of the stopping force 
should be greatly reduced as a result of the deceleration effect. Note that the applicability condition
of the linear perturbation approximation, which can be written as
\begin{equation}
\frac{Ze^2 \omega_{\rm pi}}{4\pi \varepsilon_0 m_{\rm i} v^3} =2\sqrt{ \pi} \Gamma_{\rm i}^{3/2}Z\left( \frac{v_{\rm i}}{v} \right)^3 \ll 1,
\label{linear-approximation}
\end{equation}
where $\omega_{\rm pi}=[n e^2 / (\varepsilon_0 m_{\rm i})]^{1/2}$ is the ion plasma frequency,
is violated for these parameter values as the left-hand side of the inequality~(\ref{linear-approximation}) is $\simeq 2.5$,
but it seems highly unrealistic to expect that nonlinear effects can make the deceleration correction insignificant.
For a He$^{2+}$-projectile,
the left-hand side of the inequality~(\ref{linear-approximation}) is $\simeq 0.3$,
while $\delta F/F_0 \simeq 0.24$, which is still a noticeable correction.

To conclude, we have rigorously addressed the effect of the projectile deceleration
on the stopping force. We have provided a general expression that allows investigating
this effect for any specified dielectric function [Eq.~(\ref{general-expression})].
For a cold plasma, we have derived a simple expression for the correction to the
stopping force [Eq.~(\ref{principal2})].
Assuming that the projectile is an ion,
we have found that
this correction is small when the stopping force is due to the plasma electrons,
but can be significant when the stopping force is due to the ions.

\begin{acknowledgments}
The authors thank Mikhail Pustylnik for useful discussions.
The work received funding from the European Research Council
under the European Union's Seventh Framework Programme
(FP7/2007---2013)/ERC Grant agreement 267499.
\end{acknowledgments}

\end{document}